\documentclass[prb,twocolumn,superscriptaddress,preprintnumbers,a4paper,amsmath,amssymb,showpacs,floatfix]{revtex4}
\usepackage{graphicx}
\usepackage{bm}

\newcommand{\Tc}{$T_{c}$}

\newcommand{\sqw}{$S({\bf Q},\omega)$}
\newcommand{\hbw}{$\hbar\omega$}

\begin{document}

%\preprint{}

\title{Incommensurate itinerant antiferromagnetic excitations and spin
resonance in the FeTe$_{0.6}$Se$_{0.4}$ superconductor}

\author{D.~N.~Argyriou}
\email{argyriou@helmholtz-berlin.de}
\affiliation{Helmholtz-Zentrum Berlin f\"{u}r Materialen und Energy, Glienicker Str.~100, D-14109 Berlin, Germany}

\author{A.~Hiess}
\affiliation{Institut Max von Laue-Paul Langevin, 6 rue Jules Horowitz, BP 156, F-38042, Grenoble Cedex 9, France}

\author {A. Akbari}
\affiliation {Max-Planck-Institut f\"{u}r Physik komplexer
Systeme, D-01187 Dresden, Germany}

\author {I.~Eremin}
\email{ieremin@mpipks-dresden.mpg.de}
\affiliation {Max-Planck-Institut f\"{u}r Physik komplexer
Systeme, D-01187 Dresden, Germany}
\affiliation {Institute f\"{u}r Mathematische und Theoretische
Physik, TU Braunschweig, D-38106 Braunschweig, Germany}

\author {M.M.~Korshunov}
\affiliation {Max-Planck-Institut f\"{u}r Physik komplexer
Systeme, D-01187 Dresden, Germany}
\affiliation {L.V. Kirensky Institute of Physics, Siberian Branch of Russian Academy of Sciences, 660036 Krasnoyarsk, Russia}
\altaffiliation{Present address: Department of Physics, University of Florida, Gainesville, Florida 32611, USA}

\author{Jin Hu}
\author{Bin Qian}
\author{Zhiqiang Mao}
\affiliation{Department of Physics, Tulane University, New Orleans, Louisiana 70118 USA }
\author{Yiming Qiu}
\affiliation{NIST Center for Neutron Research, National Institute of Standards
and Technology, Gaithersburg, MD 20899, USA} 
\affiliation{Dept.\ of Materials Science and Engineering, University of Maryland, College Park, MD 20742, USA}
\author{Collin Broholm}
\affiliation{Institute for Quantum Matter and Department of Physics and Astronomy, The Johns Hopkins University,Baltimore, Maryland 21218 USA  }

\author{W. Bao}
\email{wbao@ruc.edu.cn}
\affiliation{Department of Physics, Renmin University of China, Beijing 100872, China}

\date{\today}
%\preprint{Version 3.5}
\begin{abstract}

We report on inelastic neutron scattering measurements that find incommensurate itinerant like magnetic excitations in the normal state of superconducting FeTe$_{0.6}$Se$_{0.4}$ (\Tc=14K) at wave-vector 
$\mathbf{Q}_{inc}=(1/2\pm\epsilon,1/2\mp\epsilon)$ with $\epsilon$=0.09(1). In the  superconducting state only the lower energy part of the spectrum shows significant changes by the formation of a gap and a magnetic resonance that follows the dispersion of the normal state excitations.  We use a four band model to describe the Fermi surface topology of iron-based superconductors with the extended $s(\pm)$ symmetry and find that it qualitatively captures the salient features of these data. 
\end{abstract}

%%%%%%%%%%%%%%%%%%%%%%%%%
\maketitle
Magnetism is a key ingredient in the formation of Cooper pairs in unconventional high temperature superconductors\cite{chub}. A consequence of magnetism and the symmetry of the superconducting order parameter is a magnetic resonance that has been detected through inelastic neutron scattering for a wide range of magnetic superconductors ranging from the heavy fermion systems\cite{AH15,Ce122_resn,Co115_stock}, to cuprates\cite{ybco_resn,BSCCO_resn,TBCO_resn,Hayden04,John_LaBa}, and more recently in the iron based superconductors\cite{A073932,A114755,bao09a}.  In each class of materials the features of the  resonance in $\mathbf{S}({\bf Q},\omega)$ are different and are directly related to the electronic degrees of freedom, providing vital clues to the role of magnetic fluctuations in each case. We report inelastic neutron scattering measurements showing that the magnetic excitations in FeTe$_{0.6}$Se$_{0.4}$ are incommensurate and itinerant-like, and that upon entering the superconducting state only the lower energy part of the spectrum shows significant changes by the formation of a gap and a magnetic resonance that follows the dispersion of the normal state excitations.  Using a four band model that describes the Fermi surface topology of iron-based superconductors and the extended $s(\pm)$ symmetry, we can qualitatively describe the salient features of the data. The good agreement between theory and experiment found here may provide clues to a better understanding of the magnetic resonance in high-\Tc\ cuprate superconductors.

In cuprate superconductors a magnetic resonance, whose energy scales with the superconducting energy gap,  is a saddle point of a broader spectrum of magnetic excitations \cite{BSCCO_resn,TBCO_resn,Hayden04,John_LaBa}. The interpretation of the magnetic resonance and its relationship to superconductivity is complicated in the cuprates due to the occurrence of charge stripes and the pseudogap phase that lead to hotly debated models such as  quantum excitations from charge stripes\cite{John_LaBa}, or spin excitons from a particle-hole bound state of a $d$-wave superconductor\cite{resn_1,resn_2,Ismer:2007p10508}.

A far clearer picture has emerged in the iron based superconductors partly due to their more itinerant nature. It was realized early on that the nesting between hole and electron Fermi surfaces related by the antiferromagnetic wavevector {\bf Q}$_{AF} = (\pi,\pi)$ in the undoped iron superconductors, might also be responsible for unconventional superconductivity of the so-called extended $s(\pm)$-wave symmetry\cite{Mazin:1995p14594,mazin:057003,Chubukov:2008p12003,Chubukov:2008p11777,Graser:2009p12030}. The remarkable feature of this superconducting order parameter is the different sign of the superconducting gap on bands that are separated by {\bf Q}$_{AF}$  can yield a magnetic resonance in the form of a spin exciton \cite{Abanov:1999p14648,Brinckmann:1999p14649,Kao:2000p14650,Manske:2001p14651,Norman:2000p14652,Chubukov:2001p14654,Korshunov:2008p12093,Seo:2009p12214,Maier:2009p12049} at the same wavevector.

This picture suggests that the continuum of magnetic excitations will be gapped in the superconducting state up to twice the energy of the superconducting gap, $2\Delta$ and that an additional feature, the magnetic resonance will occur at an energy $\hbar\omega_{res}<2\Delta$. The latter is a consequence of both the unconventional symmetry of the superconducting order parameter which enhances the continuum at $2\Delta$, and the residual interaction between the quasiparticles that shifts the pole in the total susceptibility to lower energies. While so far the magnetic resonance has been observed in iron based superconductors at commensurate wavevectors, the itinerant origin of the magnetic fluctuations suggests that in the doped superconducting compounds, the nesting could be incommensurate.  

\begin{figure}[tb!]
\includegraphics* [scale= 0.5] {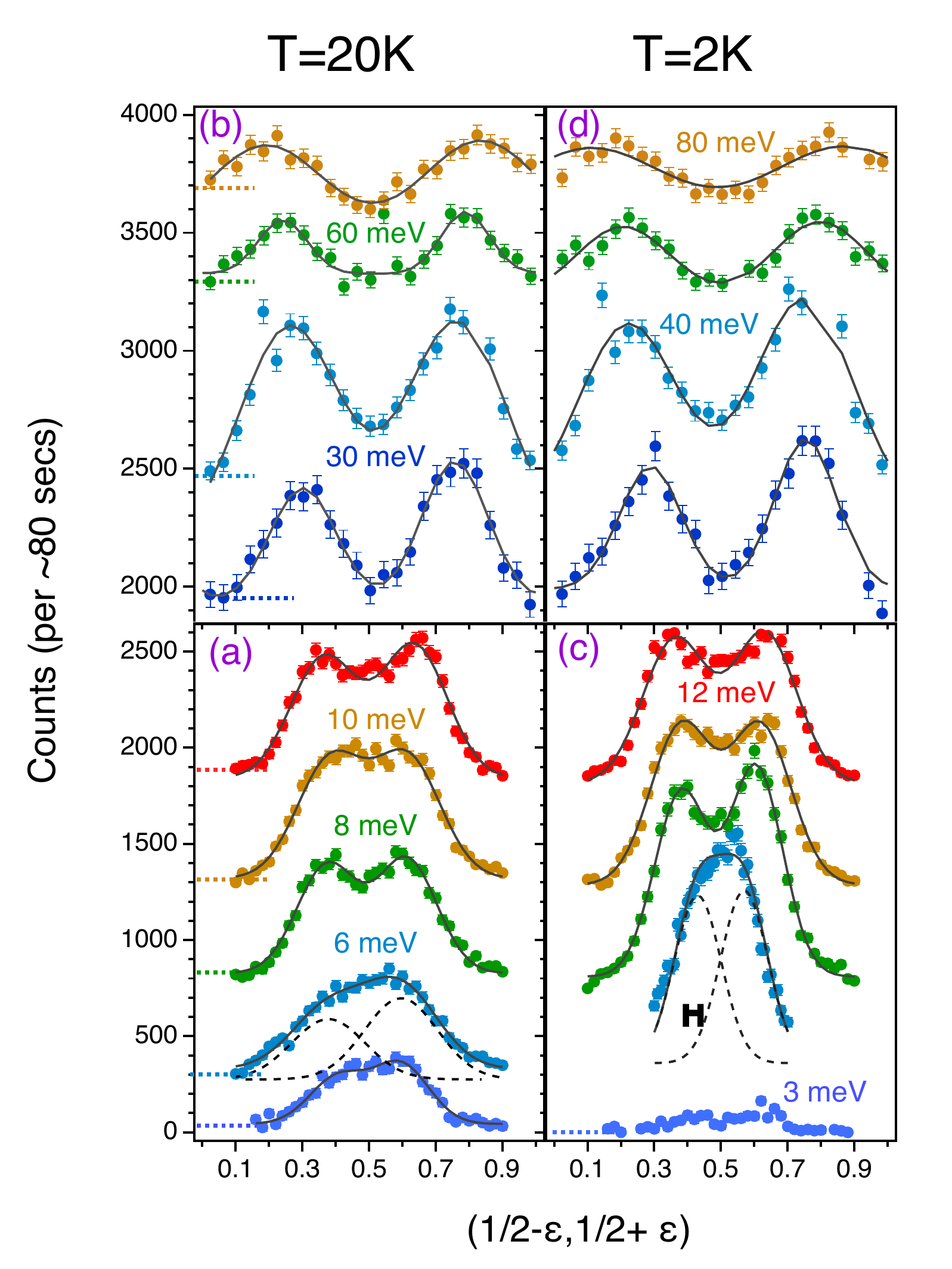}
\includegraphics* [scale= 0.6] {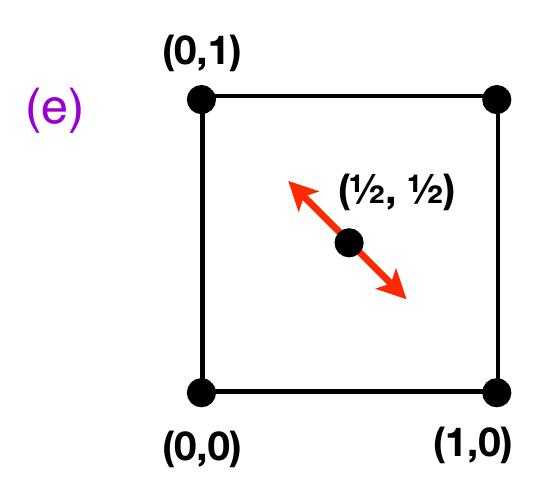}
\caption{Typical INS transverse scans through $\mathbf{Q}$=(1/2,1/2) at fix \hbw\ measured from a single crystal of FeTe$_{0.54}$Se$_{0.46}$ at 2K in the superconducting state (a,b) and at 20K in the normal state (c,d). The solid lines through the data are Gaussian fits of the magnetic excitations.  For clarity, data and fits are shifted along the $y$-axis by an arbitrary amounts. The position of the background for each scan is indicated by a horizontal line. The black horizontal bar in panel (c) represent the expected Q-resolution. (e) A map of reciprocal space used in our work. The red arrow shows the trajectory of the constant energy scans shown above. }
\label{highe}
\end{figure}

In order to understand the extended magnetic excitation spectrum in iron based superconductors we have chosen to examine the doped binary superconductor FeTe$_{0.6}$Se$_{0.4}$ with a superconducting \Tc=14K, using inelastic neutron scattering (INS). The availability of large and high-quality single crystals and the simplicity of its crystal structure (tetragonal $P4/nmm$ unit cell\cite{bao08g}) compared to other iron based superconductors make this system attractive for further examination. 
In this work we describe the measurements in momentum space in units of Q=($\pi/a$,$\pi/a$) while we omit reference to the $c$-axis direction for clarity as Q$_{z}$=0 throughout. The portion of reciprocal space probed is shown in Fig.~\ref{highe}(e), which was not explored in a previous study\cite{bao09a}.

FeTe$_{0.6}$Se$_{0.4}$ crystallize in a primitive tetragonal structure P$4/mmn$, with  $a$=3.80\AA and $c$=6.02\AA. High quality single crystal samples of this composition were prepared as described previously\cite{bao09a}. Inelastic neutron scattering measurements were collected using the triple axis spectrometer IN8 operated by the Institut Laue-Langevin (ILL), with the sample mounted with the $c-$axis vertical giving us access to spin excitations within the basal $ab$-plane. The neutron optics were set to focus on a virtual source of 30 mm. Measurements were made with the (002) reflection of a pyrolytic graphite monochromator and analyzer with  an open/open/open/open configuration.  A graphite filter was placed after the sample position. Diaphragms were placed before and after the sample. Cooling of the sample was achieved by a standard orange cryostat. Data were collected around the (1/2,1/2,0) and (3/2,3/2,0) Brillouin zones using a fixed final wave vector $k_{f}$ of either 2.66 or 4.1 \AA$^{-1}$.

\begin{figure}[t!]
\includegraphics* [scale= 0.35] {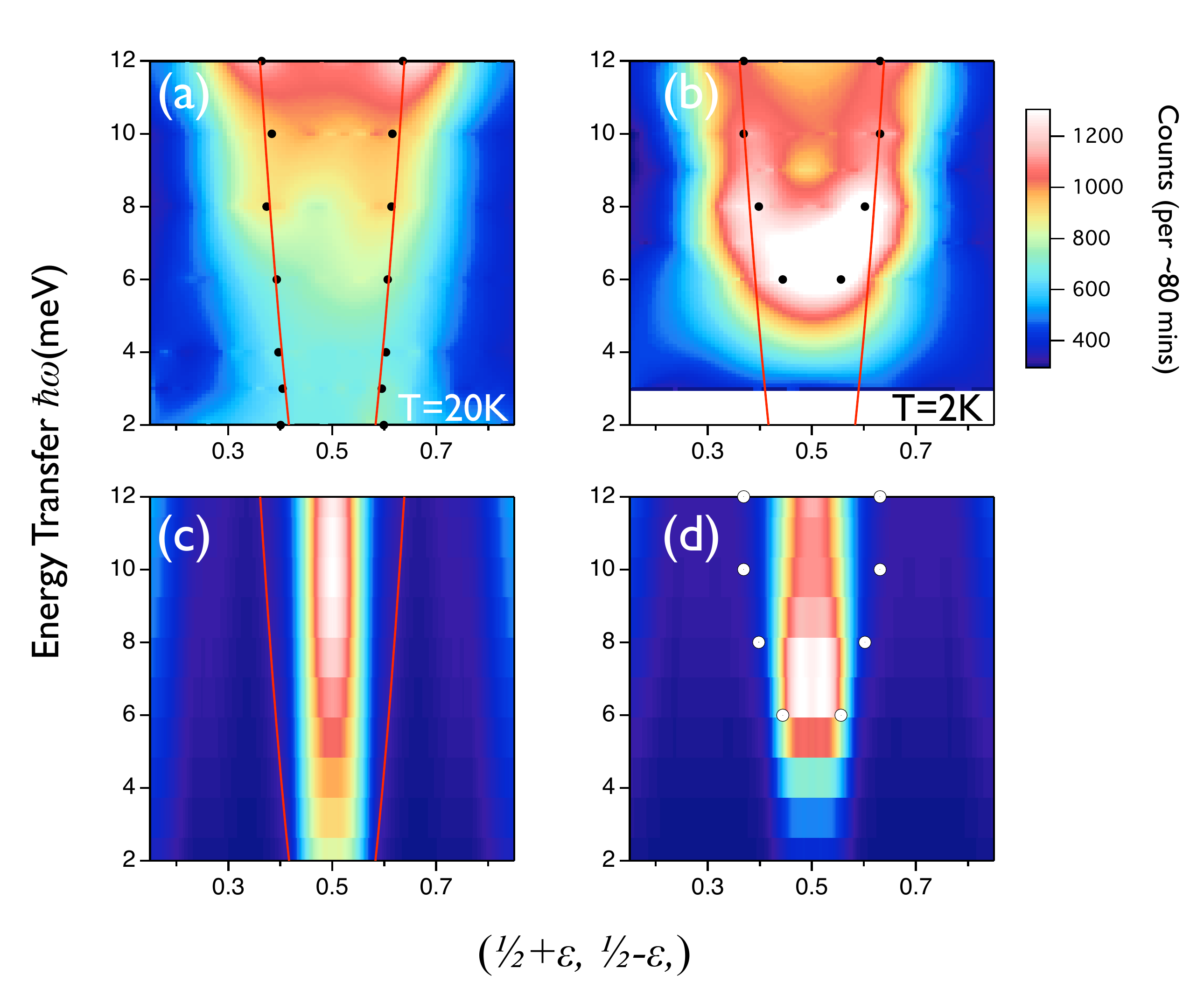}
\caption{ Color maps of the transverse constant energy scans along (1/2+$\epsilon$,1/2-$\epsilon$,0) for energy transfers between 2 and 12 meV measured (a) at 20K in the normal state and (b) at 2K in the superconducting state. Here the measured intensity is depicted as color. Black points indicate the center of Gaussian peaks fitted to the experimental data while the red lines illustrate the dispersions of these peak centers. In (b) black points represent centers of Gaussian from the substation shown in Fig.~\ref{res2}(b).  The dispersion of the normal state is reproduced by red lines  in the superconducting state to illustrate that the the higher energy excitations have not changed their position below \Tc.  RPA calculations depicting the magnetic excitations of the normal and superconducting state based on the $s(\pm)$ model of the superconducting order parameter are shown for the normal state in panel (c) and superconducting state in panel (d).  In (c) we over plot the experimental paramagnon dispersion while in (d) we show the measurement incommensurate magnetic resonance excitations as white circles. The RPA calculations shown here were convoluted with the instrument resolution function represented as a  Gaussian with FWHM of 1meV in \hbw\ and 0.06 r.l.u. in $\mathbf{Q}$. The raw calculation is shown in the Appenix.   }
\label{lowe}
\end{figure}

\begin{figure}[t!]
\includegraphics* [scale= 0.32] {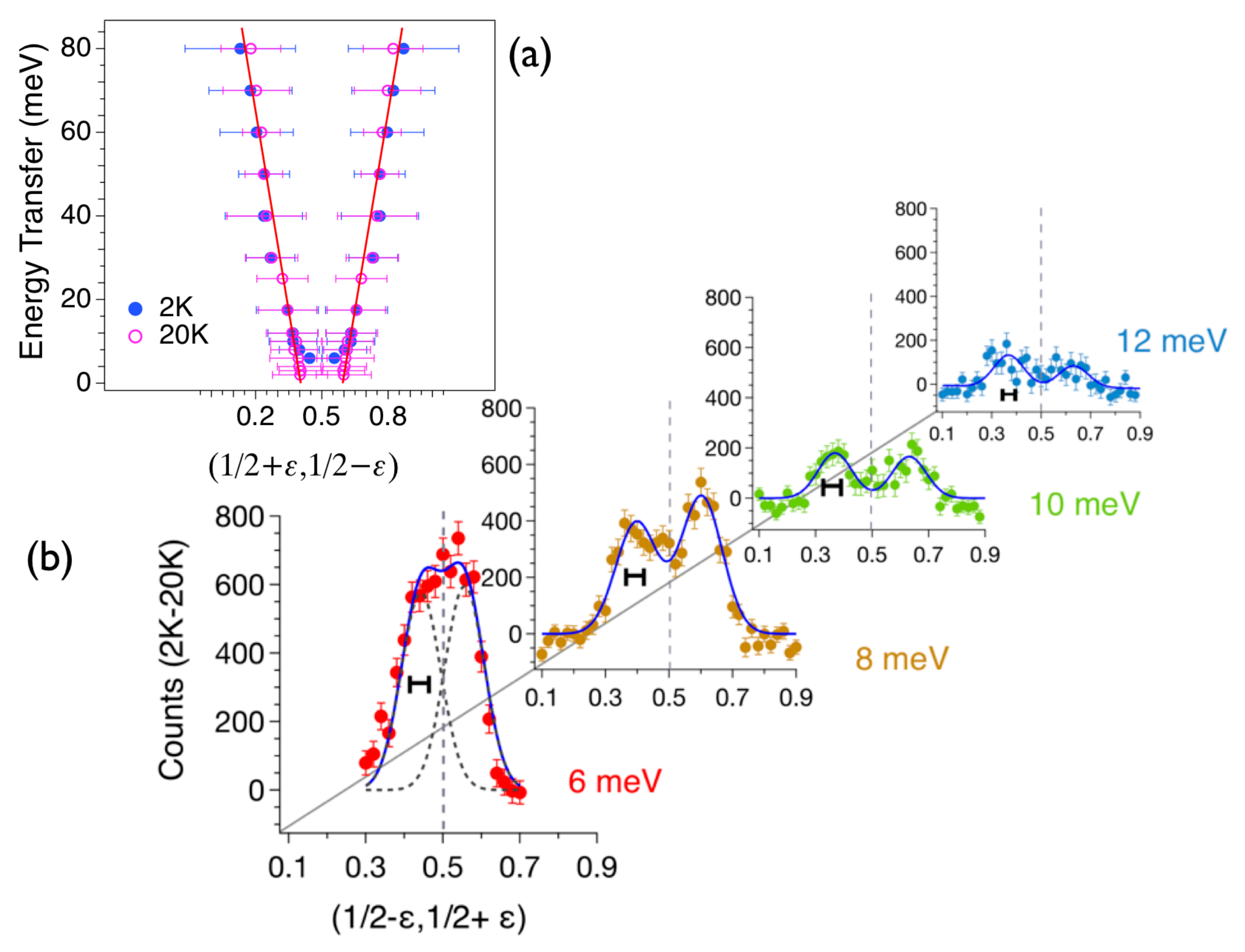}
\caption{(a) Experimental dispersions of the magnetic excitations of FeTe$_{0.6}$Se$_{0.4}$ at 20K and 2K determined by fitting Gaussian peaks to the measured excitations. A linear fit to the 20K data is illustrated as a red lines. The horizontal errors bars indicate the FWHM of the excitation determined from a fit of a Gaussian to the data. (b) Difference transverse INS scans through $\mathbf{Q}$=(1/2,1/2) at various energy transfers between 6 and 12 meV. Here in order to subtract the paramagnon scattering in the superconducting state and to obtain only the magnetic resonant contribution we subtract the data measured at 20K from the data measured at 2K. Both sets of data are shown in Fig. 1. The difference is modeled using two gaussians centered on incommensurate positions and the fit is shown as a continues line through the data. For the 6 meV data we shown the individual peaks as dashed lines. }
\label{res2}
\end{figure}

In the normal state we find steeply dispersive magnetic excitations that we have measured up to 80 meV, Fig.~\ref{highe}(a,b).  At higher energies, two excitations are clearly resolved in the transverse scans.
At lower energies (2-6 meV), the two peaks merge to a broad response, far wider than the resolution function ($\Delta$Q$\sim$~0.06 r.l.u. measured on the (200) Bragg peak).   These lower energy data at 20K are best modeled with two excitations as shown for example for the 6 meV data in Fig.~\ref{highe}(a).
In Fig.~\ref{lowe}(a) we show the lower portion of these transverse scans in the normal state as a color map where in addition we plot as black points the ${\bf Q}$-position of the excitations determined by fitting Gaussian peaks to the data of Fig.~\ref{highe}. The dispersion of these excitations is linear up to 80 meV as shown in Fig.\ref{res2}(a), and the points for the lower energy excitations fall on the same line confirming our above analysis. Extrapolating the centers of these excitations to zero energy, red lines in Fig.~\ref{lowe}(a), it is clear that they do not converge to the commensurate {\bf Q}$= (1/2,1/2)$ wavevector but to the incommensurate positions of $\mathbf{Q}_{inc}=(1/2\pm\epsilon,1/2\mp\epsilon)$ with $\epsilon$=0.09(1). We note that for the FeTe end-compound with lower excess Fe, static antiferromagnetism appears at Q=(1/2,0)\cite{bao08g}.

Longitudinal scans (not shown) between 2 and 80 meV through these spin excitations show that they are well defined isolated maxima and not parts of spin-wave cones as found in insulators, a conclusion concurred also in other recent INS measurements \cite{Lumsden:2009p11239}. Therefore, we argue that magnetic excitations in the normal state of the superconductor FeTe$_{0.6}$Se$_{0.4}$ consist of a single-lode excitation continuum from each incommensurate $\mathbf{Q}_{inc}$. This behavior in \sqw\ is reminiscent of Fincher-Burke excitations reported for many itinerant magnetic materials that are close to an antiferromagnetic instability\cite{Cr_rev,bao96a,bao06a,TiV2O3} and  indicates that an itinerant, rather than a local moment picture is relevant for this superconductor.  

Cooling below \Tc\ we find substantial changes to the lower energy magnetic excitations, while our measurements indicate that the higher energy excitations remain largely unchanged when compared to the normal state measurements, see Fig.~\ref{highe}(b,d). \footnote{We also note that the compositional dependence of these higher energy excitations appears also to be weak as shown in reference \cite{Lumsden:2009p11239}.} The most pronounced changes are in the lower energy part of the magnetic excitation spectrum where the opening of a spin gap and the development of the magnetic resonance is observed as shown in Fig.~\ref{lowe}(b). Here a spin gap opens below \hbw=4 meV, while there is an enhancement of the spectral weight above the spin gap that peaks at \hbw$_{res}$=6 meV.  At higher energies the spectral weight returns to approximately the same values as the data found in the normal state, see Fig.~\ref{highe}. Constant energy scans at the commensurate position $\mathbf{Q}$=(1/2,1/2) between these dispersions also reflect these changes as shown in reference \onlinecite{bao09a}.

Although the normal state magnetic excitations extend from incommensurate points it is important to establish  if the magnetic resonance itself is centered on the commensurate wavevector $\mathbf{Q}$=(1/2,1/2) or if it consists of two excitations originating from the two incommensurate wavevectors (at $\mathbf{Q}_{inc}$) found in the normal state.  For a more detailed analysis of the Q-scans across the resonance we subtract the 20K from the 2K data in order to obtain only the magnetic scattering contributing to the resonance (see below). These data are shown in Fig.~\ref{res2}(b), while the centers of the resonant excitations are plotted in Fig.~\ref{lowe}(b). Between \hbw=7 and 12 meV, the data shows well separated resonant excitations that follow the normal state dispersions, while their  intensity decreases with energy transfer giving only a minor superconducting enhancement at 12 meV. At the resonance, however,  $\hbar\omega_{res}=6$ meV, a ``flat top'' excitation is evident that can be modeled with two incommensurate peaks at $\mathbf{Q}_{res}=(\frac{1}{2}\pm\epsilon,\frac{1}{2}\mp\epsilon)$ with $\epsilon$=0.05(1), a value that departs slightly from the one found in the normal state.  
 
The salient point of these measurements is that in the normal state the magnetic excitations show a behavior consistent with an itinerant antiferromagnet. The onset of superconductivity affects only the lower energy magnetic excitations and spectral weight is shifted upwards to follow the normal-state incommensurate wave-vector dependence except at the spin resonance energy.  

%%%%%%%%%%%%%%%%%%%%%%%%%%%%%%%%%%%
%%  Theory part
%%%%%%%%%%%%%%%%%%%%%%%%%%%%%%%%%%%
Our INS results are qualitatively consistent with the extended s($\pm$) model of the superconducting order parameter as we indeed observe the opening of a spin gap and the enhanced peak just above the gap as expected.  However it is less clear if the magnetic resonance  and its spectral weight in \sqw\ is consistent with this model.  To clarify this we compute the magnetic susceptibility within the four-band model used previously to describe the Fermi surface topology in iron-based superconductors\cite{Korshunov:2008p12093} and use the conventional random phase approximation (RPA) which describes the enhancement of the spin response of a metal in presence of the moderately strong (interband and intraband) repulsive interactions. In Fig.~\ref{lowe}(c,d)  we show the results for the total {\em physical} RPA susceptibility,
$\chi_{RPA} ({\bf q},{\rm i} \omega_m)=\sum_{i,j}\chi^{i,j}_{RPA}({\bf q},$i$\omega_{m})$,
calculated for the Fermi surface topology that is consistent with ARPES results on Fe$_{1.03}$Te$_{0.7}$Se$_{0.3}$ \cite{Nakayama:2009p14429} as a function of the momentum along the transverse direction (1/2+$\epsilon$,1/2-$\epsilon$) and \hbw\ in the normal (c) and superconducting (d) states (details of these calculations are presented in the Supplementary information).  In order to simulate the experimental resolution we have convoluted the results by the experimental Gaussian with $\Delta {\bf q}=0.06$r.l.u. and $\Delta \omega = 1$meV and the original calculated figures are shown in the Appendix. Due to the nesting condition \footnote{{\it i.e.} $\varepsilon_{\bf k}^{\alpha}=-\varepsilon_{{\bf k+Q}_{AF}}^{\beta}$} of one of the hole-bands ($\alpha$) and the two electron-bands ($\beta$) (see Fig.~\ref{disp} in the Appendix) we find that in the normal state the spin response shows antiferromagnetic excitations centered on an incommensurate transverse wavevector with $\epsilon$=0.03. These low-energy magnetic excitations in the metals, often called \emph{paramagnons}, result from the proximity of the Fermi surface to a magnetic instability and in our model we find that they disperse linearly from the incommensurate wavevectors.
While the RPA theory indicates two peaks for each incommensurate wave vector (counter propagating linearly dispersive modes), only a single peak is observed experimentally dispersing from each $\mathbf{Q}_{inc}~$ wave vector, suggesting stronger electronic damping than the current RPA treatment, as demonstrated previously by Moriya et al.\cite{bao96a}.

In the superconducting state, the magnetic susceptibility changes due to opening of the superconducting gap and the corresponding change of the quasiparticle excitations. Here the imaginary part of the bare interband susceptibility (i.e. without
electron-electron interaction that yields RPA) is gapped for small
frequencies and at the antiferromagnetic wavevector ${\bf Q}_{AF}$, the value of this gap is determined as $\Omega_c = \min \left( |\Delta_{\bf k}| + |\Delta_{{ \bf
k+Q}_{AF}}| \right) \sim 2\Delta_0 \sim 8$meV \cite{Nakayama:2009p14429} where $\Delta_0$ is the superconducting gap. Furthermore, above $\Omega_c$ the imaginary part of the bare interband susceptibility shows the discontinuous jump implied
by the Bogolyubov coherence factor ($1+\frac{\Delta_{\bf k} \Delta_{\bf
k+Q_{AF}}}{|\Delta_{\bf k}| |\Delta_{\bf k+Q_{AF}}|}$) and the condition
$\Delta_{\bf k} = - \Delta_{{\bf k+Q}_{AF}}$. The latter is true for the
$s^{\pm}$-wave symmetry of the superconducting order parameter.
Correspondingly, the
real part of the bare susceptibility is positive, diverges logarithmically
at $\Omega_c =2\Delta_0$, and scales as $\omega^2$
at small frequencies\cite{Korshunov:2008p12093}. Then, switching on the
electron-electron interactions, we find within RPA that  for any positive
value of the interband interaction the total (RPA) spin susceptibility
acquires a pole or \emph{spin exciton} below
$\Omega_c$ with infinitely small damping. Therefore, other than the gapping of the spin spectrum,  the most salient
point of our analysis is that it predicts a spin exciton at $\hbar
\omega_{res} \approx 6$meV in the superconducting state that is located at
the same incommensurate wavevectors as paramagnons in the normal state, a
feature that is clearly found in our INS data. 
We stress here that the spin exciton is a property of the superconducting
state and requires opposite sign of the superconducting order at the
corresponding wave vector, {\it i.e.} $\Delta_{\bf k} = - \Delta_{{\bf
k+Q}_{AF}}$. We remark that the actual spin gap in the neutron
spectrum is smaller than $2\Delta_0$ which is again a consequence of the
spin exciton formation in the superconducting state. 

Our itinerant RPA like description is obviously not sufficiently accurate to describe all the detailed features of the spin excitation spectrum. For example, the value for the incommensurability $\epsilon$ is smaller than the experimentally observed and the dispersion of the paramagnons are steeper than observed as shown in Fig~\ref{lowe}(c).  Nevertheless, our treatment captures the key points of the data, these being the incommensurate paramagnon excitations, the emergence of the spin gap and the superposition of the exciton in the superconducting state onto the paramagnon scattering. 

At this point it is tempting to compare the structure in  \sqw\ of the magnetic resonance between the iron based superconductors and the cuprates. We argue that there are two basic differences.  First, in the iron based compounds the superconducting gap is almost constant at each of the Fermi surfaces, though the magnitude of the gap may differ on each of the pockets. Therefore,  the resonance is not expected to show any additional features like a downward dispersion, as found in the cuprates. Secondly, in the iron based superconductors the main electronic interactions are local (weakly momentum-dependent) and smaller than the bandwidth. The resonance in the superconducting state is an additional feature to the initial structure of the magnetic susceptibility in the normal state, determined by the nesting the Fermi surface pockets. Once the nesting is incommensurate the resonance also shifts to the incommensurate wavevectors as we demonstrate in our measurements. To confirm this we have checked that if the nesting is restored  the resonance returns to be fully commensurate. This is in contrast to the cuprates where the Fermi surface is far from being nested and the maximum intensity at $(\pi,\pi)$ of the spin response is a consequence of the momentum-dependent strong superexchange interaction.

We note that there are intriguing similarities between this iron superconductor and the rare-earth- and actinide-based superconductors  UPd$_{2}$Al$_{3}$\cite{U123_resn,upa_98,U123_2001}, CeCu$_{2}$Si$_{2}$\cite{ Ce122_resn} and CeCoIn$_{5}$ \cite{Co115_stock}. In all three compounds the magnetisation dynamics in the normal state are modulated in \sqw. In the superconducting state (i) the low energy response is suppressed by the formation of a gap, (ii) a 'resonance' is formed at energies related to the superconducting gap energy and (iii) the higher energy dynamic response remains unchanged. In this iron superconductor \Tc\ is much higher than the rare-earth- and actinide-based superconductors making this behavior far clearer to observe.  

To summarize, we find using inelastic neutron scattering that in the FeTe$_{0.6}$Se$_{0.4}$ superconductor the magnetic resonance appears superimposed at the incommensurate wavevector as paramagnon excitations found in the normal state. Our simple RPA model of the magnetic excitations on the basis of the $s(\pm)$ symmetry of the superconducting order parameters can qualitatively reproduce the main features of the neutron scattering data.  The agreement  supports an itinerant character of the spin excitations in iron-based systems and a spin spectrum determined by the single poles of Im$\chi$ with no extra contributions from localized moments. The success of this model stands in sharp contrast to the uncertainties still faced to fully understand the magnetic resonance in the cuprates.

\acknowledgements
Work at Tulane was supported by the NSF under grant DMR-0645305 (for materials) and the DOE under DE-FG02-07ER46358 (for graduate students). Work at JHU was funded by the DoE under DE-FG02-08ER46544. DNA benefited from helpful discussions with Jan Zaanen and Alan Goldman.  MMK is grateful to P.J. Hirschfeld for useful discussions and acknowledges support from RFBR (Grant N 09-02-00127) and OFN RAS program on ``Strong electronic correlations

\section{Apendix}

\subsection{Theoretical Calculations}

\begin{figure}[b!]
\includegraphics* [scale= 0.4] {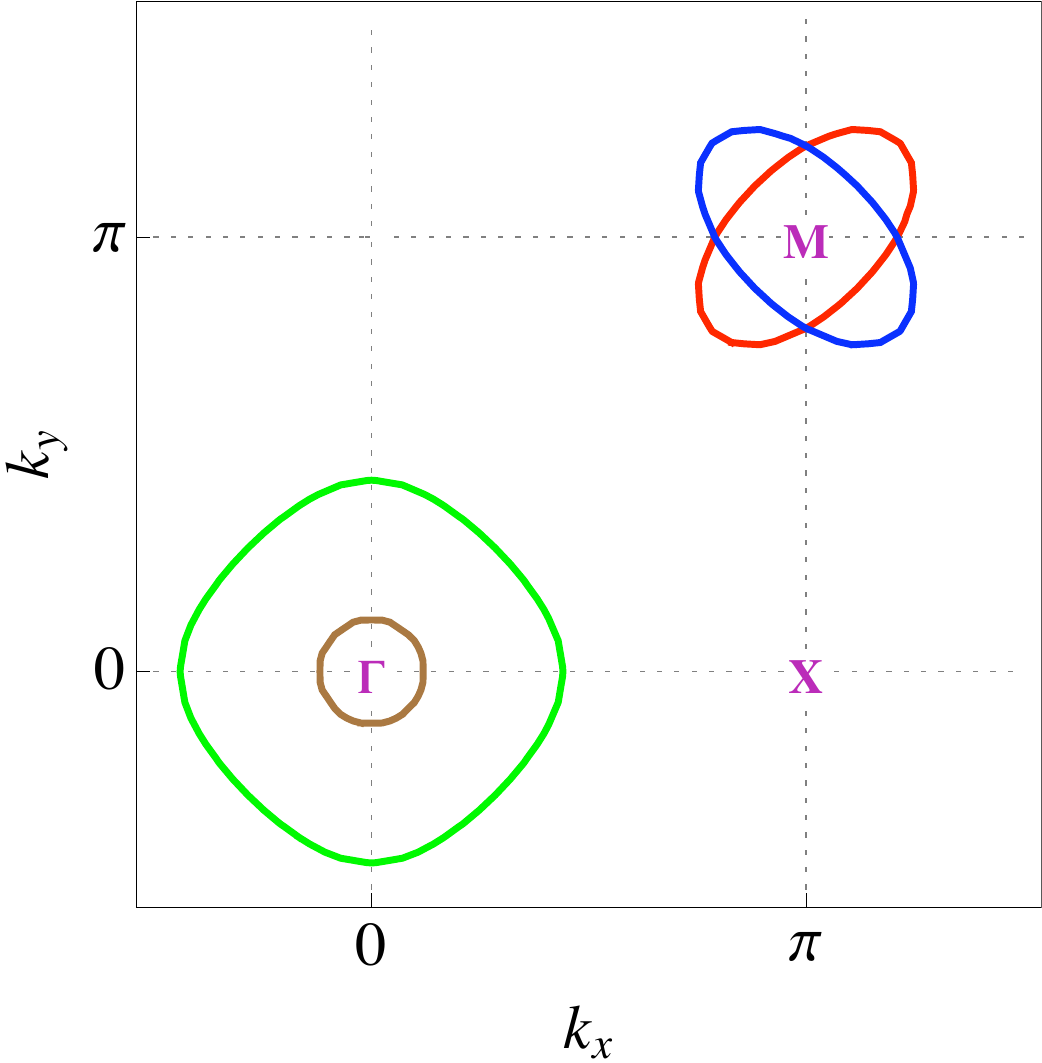}
\caption{Calculated Fermi surface topology for the four-band tight-binding Hamiltonian, Eq.(1). The circular pockets around the $\Gamma-$point refer to the hole $\alpha$-bands, and ellipses arise due to electron $\beta$-bands and are centered around the $M$-point.}
\label{disp}
\end{figure}

The band structure calculations as well as the ARPES measurements show that
the Fermi surface of iron-based superconductors consists of two hole (h) pockets
centered around the $\Gamma=(0,0)$ point and two electron (e) pockets centered
around the $M=(\pi,\pi)$ point of the {\it folded} Brillouin Zone (BZ)
\cite{mazin:057003}. We model the resulting band structure by using
the following single-electron model Hamiltonian
\begin{eqnarray}
H_0 = - \sum\limits_{{\bf k},\alpha ,\sigma } {{\epsilon^i} n_{{\bf
k} i \sigma } } - \sum\limits_{{\bf k}, i, \sigma}  t_{{\bf k}}^{i}
d_{{\bf k} i \sigma }^\dag d_{{\bf k} i \sigma}, \label{eq:H0}
\end{eqnarray}
where $i=\alpha_1,\alpha_2,\beta_1, \beta_2$ refer to the band
indices, $\epsilon^i$ are the on-site single-electron energies,
$t_{{\bf k}}^{\alpha_{1},\alpha_{2}} = t^{\alpha_{1},\alpha_{2}}_1
\left(\cos k_x+\cos k_y \right)+ t^{\alpha_{1},\alpha_{2}} _2 \cos
k_x \cos k_y$ is the electronic dispersion that yields hole pockets
centered around the $\Gamma$ point, and $t_{{\bf
k}}^{\beta_{1},\beta_{2}} = t^{\beta_{1},\beta_{2}}_1 \left(\cos
k_x+\cos k_y \right)+ t^{\beta_{1},\beta_{2}}_2 \cos \frac {k_x}{2}
\cos \frac{k_y}{2}$ is the dispersion that results in the electron
pockets around the $M$ point. This model has been used previously to fit the available ARPES data\cite{Nakayama:2009p14429} in optimally doped Ba$_{0.6}$K$_{0.4}$Fe$_2$As$_2$. It has been found recently\cite{Nakayama:2009p14429}that the Fermi surface and the
corresponding electronic structure of Fe$_{1.03}$Te$_{0.7}$Se$_{0.3}$ is very similar to the 122 compounds except for the different level of nesting. To account for this change we use the following parameters [$(\epsilon^i,
t_1^i, t_2^i)$] $(-0.26, 0.16, 0.052)$ and
$(-0.18, 0.16, 0.052)$ for the $\alpha_1$ and $\alpha_2$ bands,
respectively, and $(0.68,0.38, 0.8 )$  and $(0.68,0.38, -0.8)$  for
the $\beta_1$ and $\beta_2$ bands, correspondingly (all values are
in eV). This parametrization is almost the same as the one for Ba$_{0.6}$K$_{0.4}$Fe$_2$As$_2$\cite{Nakayama:2009p14429} though the slightly different on-site energies and different parametrization of the $\alpha_2$ band have been used to ensure the incommensurate nesting wave vectors between $\alpha$ and $\beta$-bands.

Next we consider the one-loop contribution to the spin susceptibility that
includes the intraband and the interband contributions:
\begin{eqnarray}
\chi_0^{ij}({\bf q},{\rm i} \omega_m)&=&- \frac{T}{2N}
\sum_{{\bf k}, \omega_n} {\rm Tr}
\left[ G^i({\bf k + q}, {\rm i} \omega_n + {\rm i} \omega_m) G^j({\bf k} , {\rm i} \omega_n) \right. \nonumber \\
&+& \left. F^i({\bf k + q}, {\rm i} \omega_n + {\rm i} \omega_m) F^j({\bf k} , {\rm i} \omega_n)\right]
\label{eq:bare_chi}
\end{eqnarray}
where $i$, $j$ again refer to the different band indices. $G^i$ and $F^i$ are
the normal and anomalous (superconducting) Green functions, respectively.

For the four-band model considered here the effective interaction will consist
of the intraband and interband repulsion denoted by $U$ and $J$, respectively.
Within RPA the spin response can be written in a matrix
form:
\begin{eqnarray}
\hat{\chi}_{RPA}({\bf q},{\rm i}\omega_m)=\left[\mathbf{I}-{\bf
\Gamma} \hat{\chi}_0({\bf q},{\rm i}\omega_m)\right]^{-1}
\hat{\chi}_0({\bf q},{\rm i}\omega_m) \label{eq:chi_RPA}
\end{eqnarray}
where ${\bf I}$ is a unit matrix and $\hat{\chi}_0({\bf q},{\rm i}\omega_m)$ is
$4 \times 4$ matrix formed by the interband and intraband bare susceptibilities
determined by Eq.~(\ref{eq:bare_chi}). The vertex is given by
\begin{eqnarray}
{\bf \Gamma} = \left[\begin{array}{cccc} U & J/2 & J/2 & J/2
\\ J/2 & U & J/2 & J/2 \\ J/2 & J/2 & U & J/2 \\ J/2 & J/2 & J/2 & U
\end{array}\right],
\label{eq_coupling}
\end{eqnarray}
and we assume here $J=0.05$eV and $U \sim 0.13$eV. Note that
the value of $U$ was chosen in order to stay in the paramagnetic
phase.

\begin{figure}[ht!]
\includegraphics* [scale= 0.4] {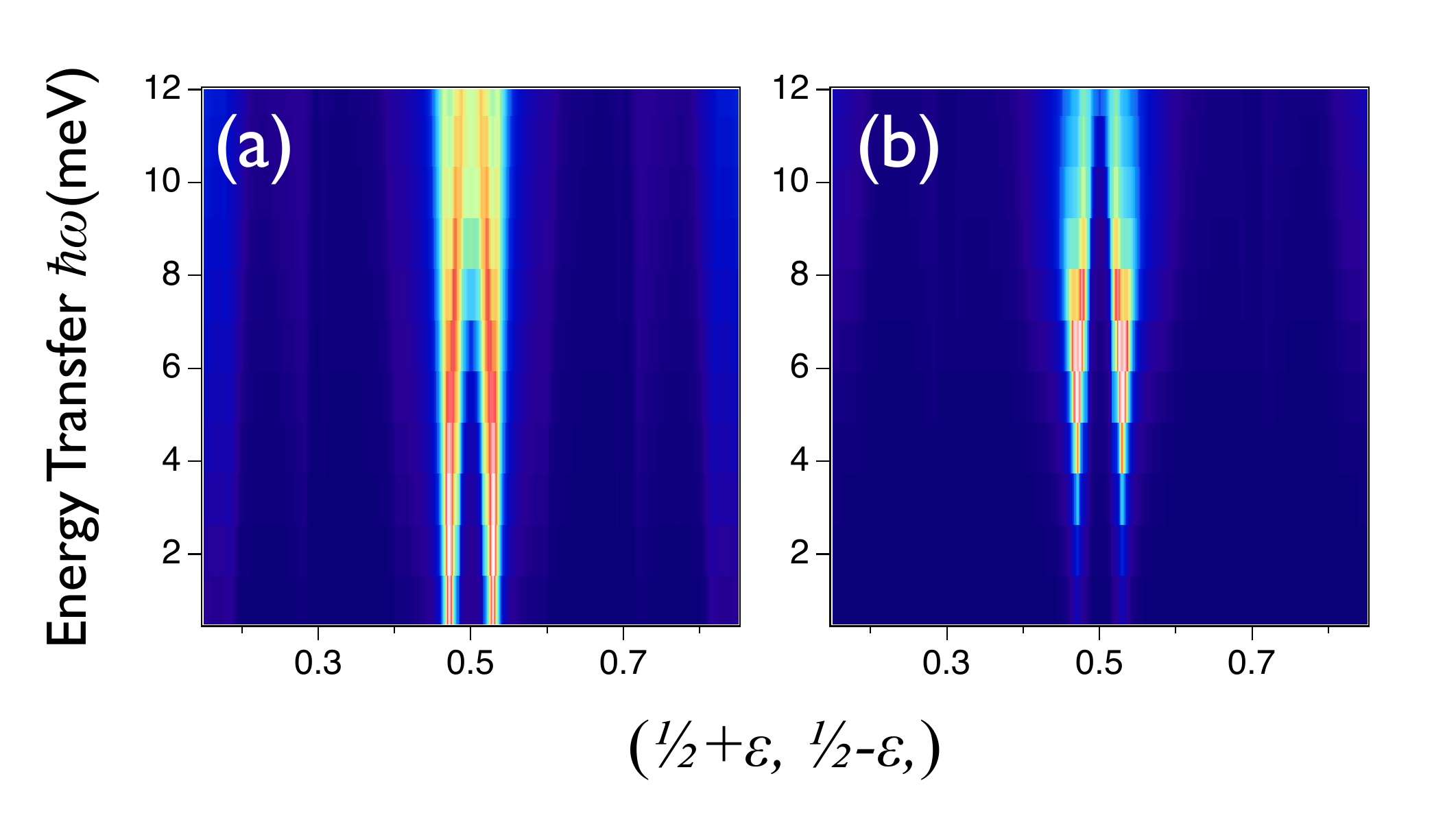}
\caption{Calculated $({\bf q}, \omega)$ mesh of the physical spin susceptibility in the normal (a) and superconducting (b) state without taking into account the resolution function.
}
\label{suscept}
\end{figure}
The results are shown in Fig.\ref{suscept} where one clearly finds the incommensurate response in both normal and superconducting state.

\bibliographystyle{h-physrev}
%\bibliography{bibliography2,for_dimitri}

\end{document}